\documentclass[structabstract]{aa}   
\usepackage{graphicx} 
\usepackage{txfonts} 
\begin{document} 
  \title{Adaptive optics near infrared integral field spectroscopy of 
      NGC~2992\thanks{based on observations collected at the European 
      Southern Observatory, Chile (074.B-9012).} 
} 
 
%   \subtitle{} 
 
\author{} 
   \author{S. Friedrich \inst{1} 
           \and R. I. Davies \inst{1} 
           \and E. K. S. Hicks \inst{1} 
	   \and H. Engel \inst{1} 
           \and F. M\"uller-S\'anchez \inst{1,2} 
           \and R. Genzel \inst{1,3} 
           \and 
           L. J. Tacconi \inst{1}  
          } 
\institute{} 
   \institute{Max-Planck-Institut f\"ur extraterrestrische Physik, Postfach 
     1312, D-85741 Garching 
         \and 
Instituto de Astrof\'isica de Canarias, E-38205 La Laguna (Tenerife), Spain 
\and 
Department of Physics, Le Conte Hall, University of  
        California, Berkeley, CA, 94720, USA 
} 
 
%             \email{c.ptolemy@hipparch.uheaven.space} 
%             \thanks{The university of heaven temporarily does not 
%                     accept e-mails} 
              
\offprints{S. Friedrich \\ \email{sfriedrich@mpe.mpg.de}}

   \date{} 
 
% \abstract{}{}{}{}{}  
% 5 {} token are mandatory 
  
  \abstract 
  % context heading (optional) 
  % {} leave it empty if necessary   
   {} 
  % aims heading (mandatory) 
   {NGC~2992 is an intermediate Seyfert 1 galaxy showing outflows on 
       kilo parsec scales which might be due either to AGN or starburst 
       activity. We therefore aim at investigating its central region for a 
       putative starburst in the past and its connection to the AGN and the outflows.} 
  % methods heading (mandatory) 
   {Observations were performed with the adaptive optics near infrared integral 
     field spectrograph SINFONI on the VLT, complemented by longslit observations 
     with ISAAC on the VLT, as well as N- and Q-band data from the Spitzer archive. 
     The spatial and spectral resolutions of the SINFONI data are 50\,pc and 
     83\,km\,s$^{-1}$, respectively. The field of view of 
     $3\arcsec\times 3\arcsec$ corresponds to 450\,pc $\times$ 450\,pc. Br$\gamma$ 
     equivalent width and line fluxes from PAHs were  
     compared to stellar population models to constrain the age of the putative 
     recent star formation. A simple geometric model of two mutually inclined 
     disks and an additional cone to describe an outflow was developed to 
     explain the observed complex velocity field in H$_2$\,1-0S(1).} 
  % results heading (mandatory) 
   {The morphologies of the Br$\gamma$ and the stellar continuum are different 
     suggesting that at least part of the Br$\gamma$ emission comes from the AGN. This 
     is confirmed by PAH emission lines at 6.2\,$\mu$m and 11.2\,$\mu$m and 
     the strength of the silicon absorption feature at 9.7\,$\mu$m, which 
     point to dominant AGN activity with a relatively minor starburst contribution.  
     We find a  
     starburst age of 40\,Myr -- 50\,Myr from Br$\gamma$ line diagnostics 
     and the radio continuum; ongoing star formation can be 
  excluded. Both the energetics and the timescales indicate that the 
  outflows are driven 
     by the AGN rather than the starburst. The 
     complex velocity field observed in H$_2$\,1-0S(1) in the central 450\,pc can 
     be explained by the superposition of the galaxy rotation and an 
     outflow. } 
  % conclusions heading (optional), leave it empty if necessary  
   {} 
 
   \keywords{galaxies: active -- galaxies: individual: NGC 2992 --  
             galaxies: Seyfert -- galaxies: starburst -- infrared: galaxies 
               } 
 
   \authorrunning{S. Friedrich et al.} 
%   \titlerunning{NGC 2992} 
   \maketitle 
% 
%________________________________________________________________ 
 
\section{Introduction} 
\object{NGC 2992}, initially classified as a Seyfert 2, was later classified as an 
intermediate Seyfert 1 galaxy on the basis of a broad H$\alpha$ component 
with no corresponding H$\beta$ component in its nuclear spectrum (Ward et 
al. \cite{ward80}). Later, Glass (\cite{glass97}) suggested that it is a hybrid 
between an intermediate Seyfert and a starburst galaxy, induced by the 
interaction with NGC 2993 to the south. Both galaxies are connected by a 
tidal tail with a projected length of 2\farcm9.     
 
NGC 2992 is highly inclined by about 70$\degr$ to our line of 
sight with a broad disturbed lane of dust in its equatorial plane.  
The redshift is 0.0077 which corresponds to a 
distance of 32.5\,Mpc and an angular scale of 150\,pc per arcsec assuming  
$H_0$=75\,km\,s$^{-1}$\,Mpc$^{-1}$. 
 
At a wavelength of 6\,cm NGC 2992 shows prominent loops to the northwest and 
southeast. This feature, designated as figure-of-8, has a size  
of about 8\arcsec\ with the long axis at a position angle (PA) of 
$-26\degr$, which is not coincident or perpendicular to the major axis 
(Ulvestad \& Wilson \cite{ulvestad84}). Chapman et 
al. (\cite{chapman00}) conclude from their IR observations that the most convincing 
interpretation for these loops is that of expanding gas bubbles which are seen 
as limb-brigthened loops as was already proposed by Wehrle \& Morris 
(\cite{wehrle88}).  
Two possible explanations are discussed for the origin 
of the loops: They may be driven by an AGN or a superwind from a 
compact starburst. X-ray observations point to the first possibility:  
Colbert et al. (\cite{colbert98}) concluded that the large scale 
outflows are driven by nonthermal jets from  
the AGN that entrain the surrounding material and heat it by shocks over kpc 
spatial scales.  
 
Heckman et al. (\cite{heckman81}) reported complex kinematics in NGC 2992 on 
the basis of longslit spectra at PA=120\degr. This work was extended by Colina 
et al. (\cite{colina87}) who find indications for an outflow within a plane 
highly inclined to the plane of the galaxy.  
M\'arquez et al. (\cite{marquez98}) measured [\ion{O}{iii}] and H$\alpha$ lines at nine 
different position angles and concluded, although from a simple kinematical 
model, that the double peaked line profiles and asymmetries can be accounted 
for by an outflow within a conical envelope or on the surface of a hollow 
cone. The high spectral resolution observations (4.9\,\AA: 3440\,\AA\ -- 
5579\,\AA, 11.9\,\AA: 5840\,\AA\ -- 9725\,\AA) of narrow line regions in NGC 
2992 of Allen et al. (\cite{allen99}) show at most positions a double 
component line profile. One component of the double peaked lines follows the 
galactic rotation curve, while the other is identified as a wind that is 
expanding out of the plane of the galaxy with a velocity of up to 
200\,km\,s$^{-1}$. These authors did not find spectral evidence for a 
significant contribution from a starburst to the outflow.  
 
Depending on the inclinations of galaxy and cone several interpretations of 
these kinematics are possible. It is generally assumed that the NW edge of the 
disk of NGC 2992 is closest to us with trailing spiral arms, and that the SE 
cone is closer and directed at us. However, an arc-shaped emission found by 
Garc\'ia-Lorenzo et al. (\cite{garcia01}), which might be associated with the 
radio loop, is red shifted in the SE, and can therefore only be consistent 
with an outflow if it is pointing away from us.   
 
In this paper, we present spatially and spectrally resolved 
  K-band observations of the  
  inner $3\arcsec\times3\arcsec$ of NGC 2992 taken with the adaptive 
  optics integral field spectrograph SINFONI.  
These are complemented by longslit observations along the major axis  
out to a radius of about 5\arcsec\ with ISAAC, and N- and Q-band spectra from 
the Spitzer archive. SINFONI data were taken with unprecedent 
  spatial resolution in the K-band which allows us to study the dynamics of gas 
  and stars in the central region in detail. Together with the spectral 
  resolution  
  of SINFONI this enables us to search for a starburst and its connection 
  to the AGN and, possibly, to the outflows. After presenting the 
  observational results in section 3, we discuss the observations in section 
  4, and show that there is evidence for a nuclear starburst in 
  NGC 2992. Finally, in section 5 we concentrate on the 
dynamics of the gas and stars and the consequences on the geometry of the 
galaxy.

\section{Observations and data reduction} 
\subsection{SINFONI} 
K-band spectra of NGC 2992 between approximately 1.95\,$\mu$m and 2.45\,$\mu$m 
were taken on 13 March 2005 at the VLT with SINFONI,  
an adaptive optics near infrared integral field spectrograph (Eisenhauer et al. 
\cite{eisenhauer03}; Bonnet et al. \cite{bonnet04}) 
with a spectral resolution of 3600. The 
pixel scale was $0\farcs05 \times 0\farcs1$ and the exposure time 600\,s for 
both object and sky frames. Object frames were interspersed with sky frames 
using the sequence O-S-O-O-S-O. A total of 12 object frames were obtained, 
resulting in a total exposure time of two hours. For a near diffraction limited 
correction the nucleus of the galaxy was used by the AO module. 
 
Data were reduced in a standard manner by subtraction sky frames, 
flatfielding, correcting bad pixels, and wavelength calibration with an arc 
lamp using the SPRED software package (Abuter et al. \cite{abuter06}), which 
also performs the reconstruction of the data cube. No spatial smoothing by a median 
filter was applied. Telluric 
correction and flux calibration were performed with the G2V star HD 
85724 (K=7,881). Residuals from the OH line emission were removed using the method 
described in Davies (\cite{davies07a}).  
 
The spatial resolution has been estimated from both the broad Br$\gamma$ and 
the non-stellar continuum (Davies \cite{davies07b}). Both methods yield 
symmetric  
PSFs with FWHMs of 0\farcs32 and 0\farcs29, respectively, corresponding to 
48\,pc and 44\,pc, respectively. The spectral resolution derived from sky 
lines is 83\,km\,s$^{-1}$. NGC 2992 is part of a (heterogeneous) sample 
of type  
1 and type 2 Seyfert galaxies, ULIRGs, and a Quasar, which were observed to 
investigate the connection between star formation and AGN. Further details on 
data reduction can be found in Davies et al. (\cite{davies07}) and Hicks et 
al. (\cite{hicks09}). 
 
\subsection{ISAAC} 
NGC 2992 was also observed on 13 January 2006 with the Infrared Spectrometer 
And Array Camera ISAAC at the VLT in order to obtain wide FOV longslit spectra 
centered on 2.1\,$\mu$m (to measure H$_2$\,1-0S(1)) and 2.29\,$\mu$m 
(to measure $^{12}$CO(2-0)). The 2\arcmin\ long and 0\farcs3 wide slit was 
aligned parallel to the morphological large scale major (PA=34.62\degr) and 
minor axes (PA=$-$55.38\degr), 
respectively. The exposure time amounts to 40 minutes per slit position. The spectra 
were reduced using standard techniques carried out by the ESO pipeline 
software.  
 
In order to determine line positions, and subsequently radial velocities of 
H$_2$\,1-0S(1), a spectrum from the outer edge of the galaxy's major or minor axes 
was extracted, the signal-to-noise determined and if necesary the spectrum 
from the adjoining pixel added until 
a S/N=3 was reached. This procedure was repeated along the slit to the 
opposite edge of the galaxy resulting in a set of 114 and 62 individual spectra along the 
major and minor axes, respectively. Br$\gamma$ (2.16\,$\mu$m) can also be 
detected in these spectra. These 2D-data sets were then analysed in a 
similar way to their 3D counterparts (see section 5).

\section{Observational results} 
 
\subsection{The spectrum of NGC 2992} 
 
   \begin{figure} 
   \centering 
   \includegraphics[width=8truecm]{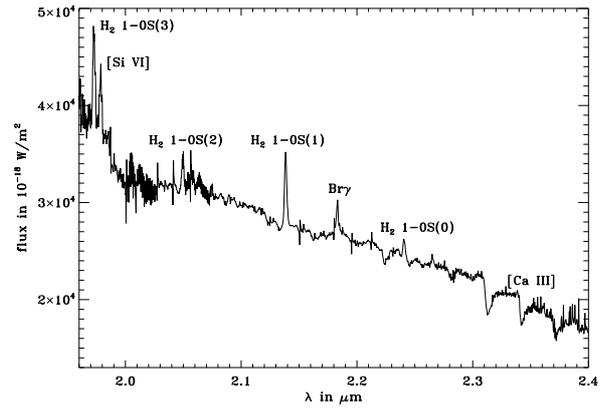} 
      \caption{SINFONI-Spectrum of NGC 2992 integrated over the whole FOV of 
        $3\arcsec\times 3\arcsec$. The most prominent lines are those of H$_2$\,1-0S(3) at 
        1.958\,$\mu$m, $\left[\ion{Si}{vi}\right]$ at 1.96\,$\mu$m, 
        H$_2$\,1-0S(1) at 2.12\,$\mu$m and Br$\gamma$ at 2.16\,$\mu$m.} 
         \label{totspec} 
   \end{figure} 
 
In Fig.\ \ref{totspec} we show the SINFONI spectrum of NGC 2992 integrated over the 
whole FOV of $3\arcsec\times 3\arcsec$ centered on the nucleus. The most 
prominent lines are  
H$_2$\,1-0S(3) at 1.958\,$\mu$m, $\left[\ion{Si}{vi}\right]$ at 1.96\,$\mu$m,  
H$_2$\,1-0S(1) at 2.12\,$\mu$m, and Br$\gamma$ at 2.16\,$\mu$m, other molecular 
hydrogen transitions and some CO bandheads are also present. The line fluxes are 
summarized in Table \ref{linflux} together with line fluxes from Gilli et 
al. (\cite{gilli00}).  
 
\begin{table}[t] 
\begin{minipage}[t]{\columnwidth} 
\caption{Measured line fluxes for NGC 2992 in a field of view of 
  $1\farcs5\times1\farcs5$ centered on the nucleus}        % title of Table 
\label{linflux}      % is used to refer this table in the text 
\centering                          % used for centering table 
\renewcommand{\footnoterule}{}  % to avoid a line before footnotes 
\begin{tabular}{l r r r}        % centered columns (4 columns) 
\hline\hline                 % inserts double horizontal lines 
%Line & $\lambda$\footnote{wavelengths are given in the rest frame} & 
%Flux\footnote{this paper} & Flux\footnote{from Gilli et al. (\cite{gilli00})} \\      
 Line &\ \hfill$\lambda$\footnote{wavelengths are given in the 
  rest frame}\hfill\ &  
\ \hfill Flux\footnote{this paper}\hfill\ &\ \hfill Flux\footnote{from Gilli 
  et al. (\cite{gilli00}), aperture size 
  $1\,\arcsec\times2\,\arcsec$}\hfill\ \\       
  &\ \hfill$$($\mu$m)$$\hfill\ & (10$^{-18}$W\,m$^{-2}$) & 
(10$^{-18}$W\,m$^{-2}$) \\  
\hline                        % inserts single horizontal line 
 H$_2$\,1-0S(3)& 1.9576& $11\pm0.2$ & 5 \\ 
 $\left[\ion{Si}{vi}\right]$& 1.9634&$13\pm1.6$& 4 \\ 
 H$_2$\,1-0S(2)& 2.0338& $3.9\pm0.7$ & 2.4 \\ 
% \ion{He}{i}& 2.0587&& 7.\\ 
 H$_2$\,1-0S(1)& 2.1218& $9.4\pm0.2$ & 2.5 \\ 
 Br$\gamma$ & 2.1661& $14.3\pm3.5$ & 11.6\\ 
 Br$\gamma$ (narrow)&2.1661&$4.8\pm1 $&--\\ 
 H$_2$\,1-0S(0)& 2.2233& $2.4\pm1.2$& --\\ 
 %\ion{Ca}{viii}& 2.3213& $1\pm0.02$& --\\ 
\hline                                   %inserts single line 
\end{tabular} 
\end{minipage} 
\end{table} 
   
   \begin{figure*} 
   \centering 
   \includegraphics[width=11.3truecm]{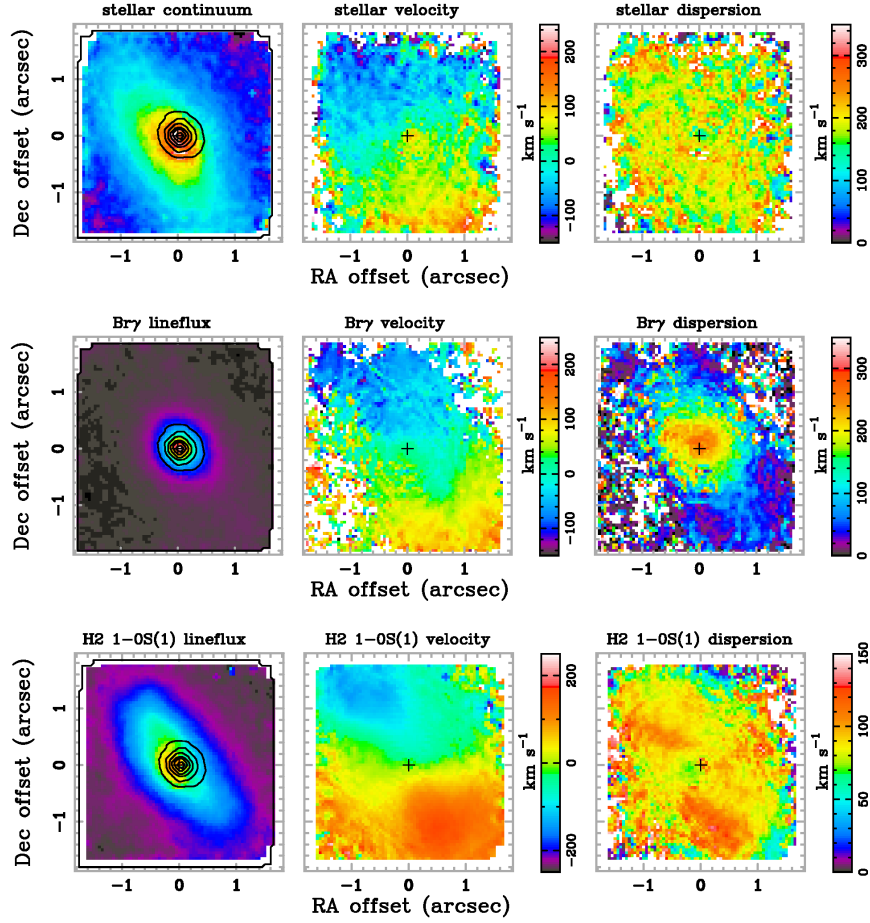} 
      \caption{Continuum, velocity, and dispersion maps of NGC 2992 (left to 
        right) of stars, Br$\gamma$, and H$_2$\,1-0S(1) (top to 
        bottom). 1\arcsec\ corresponds to 150 pc. The contour lines of the 
        Br$\gamma$ continuum are overlayed on all continuum maps and a cross 
        indicates the center of each map. North is up and east to the left.} 
         \label{kine} 
   \end{figure*} 
 
   \begin{figure*} 
   \centering 
   \includegraphics[width=8truecm]{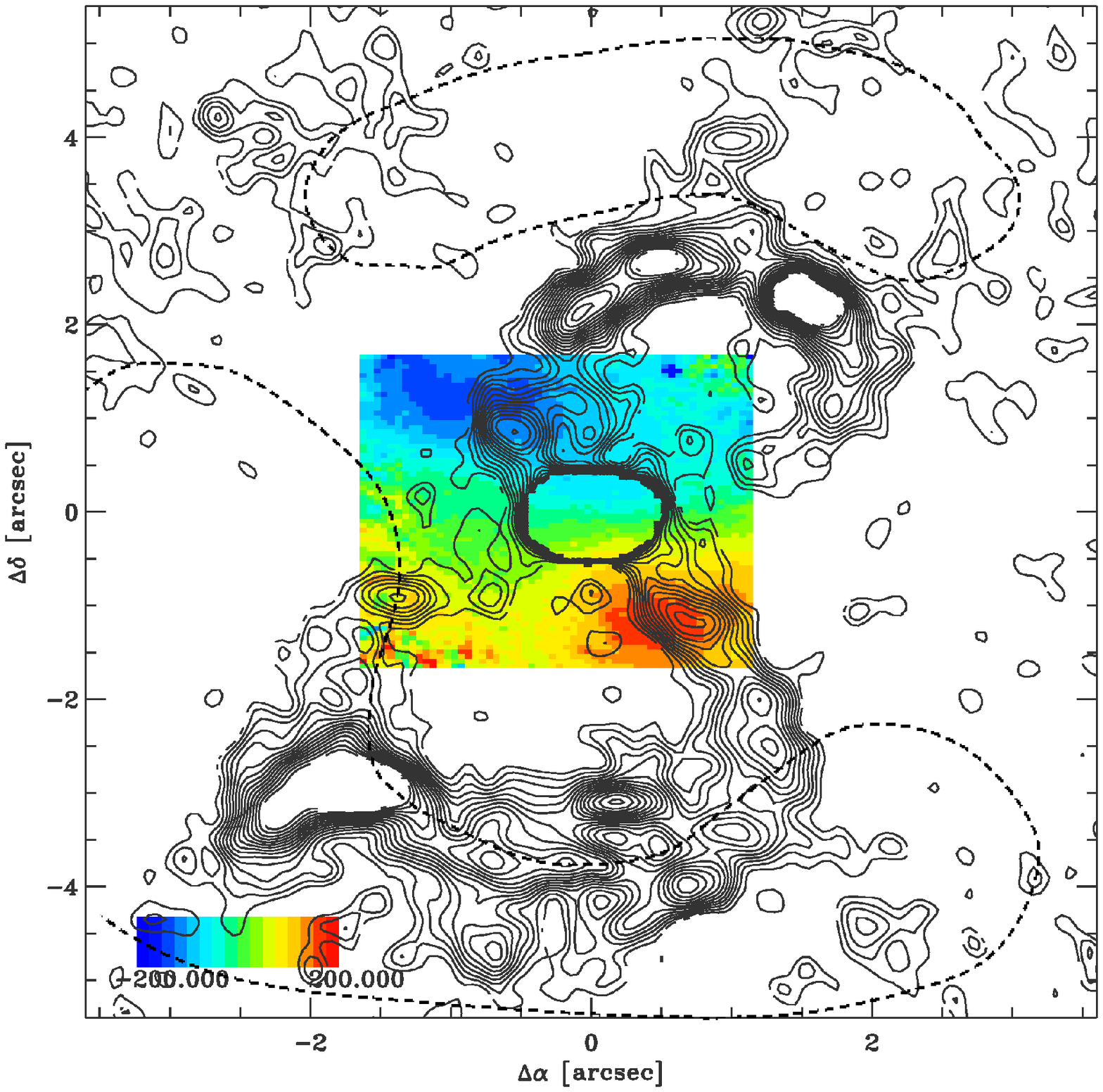} 
   \includegraphics[width=8truecm]{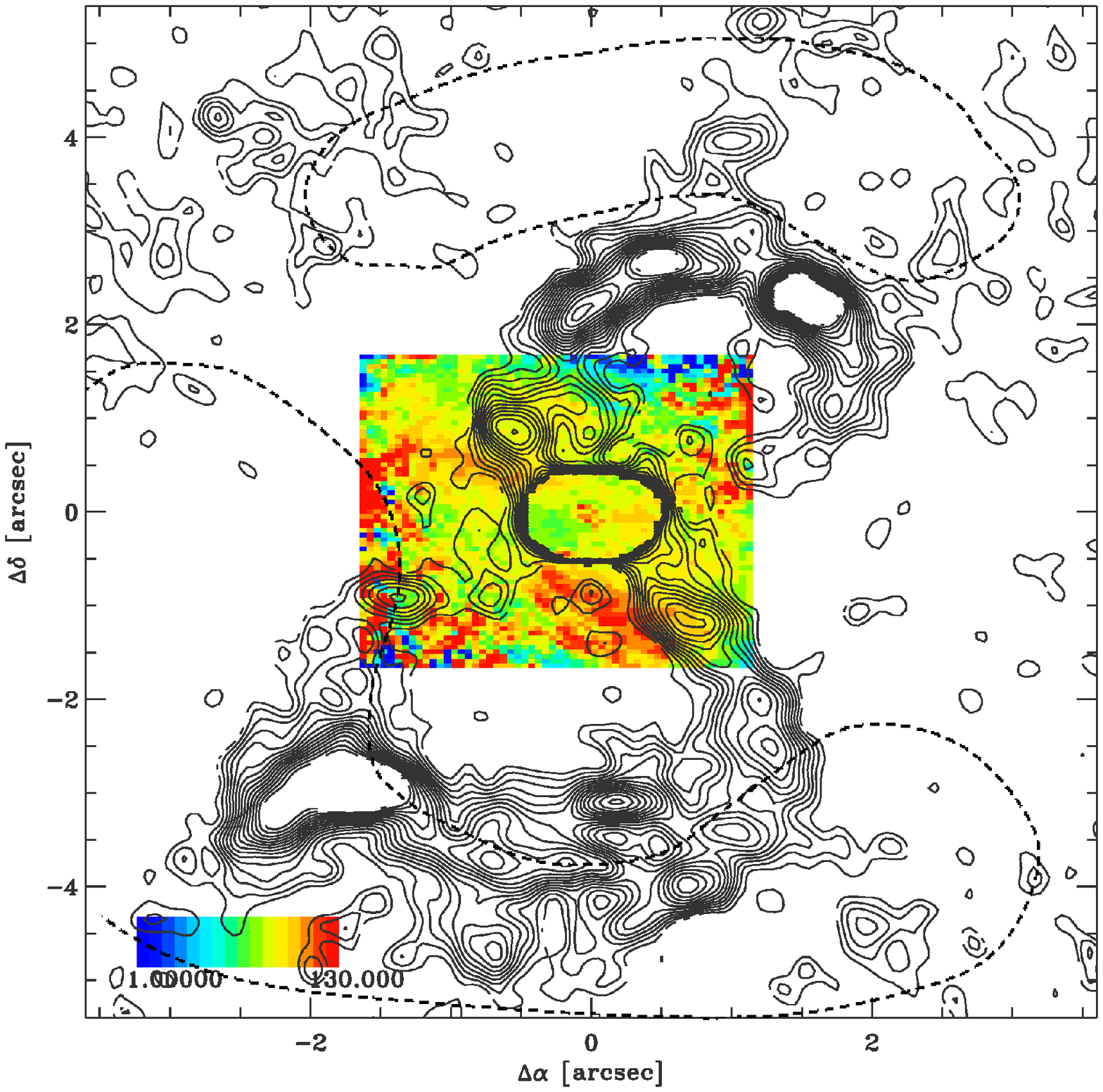} 
      \caption{Velocity (left) and velocity dispersion maps (right) of 
        H$_2$\,1-0S(1) with the contours of the radio flux  
        at 8.46\,GHz (Leipski et al. \cite{leipski06}, courtesy Heino Falcke) 
        and  
        the rough positions of the redshifted and blueshifted velocity 
        residuals of [\ion{O}{iii}] (dashed 
        lines at bottom and top) found by Garc\'ia-Lorenzo et 
        al. (\cite{garcia01}) overlayed. North is up and east to the left.} 
         \label{radio} 
   \end{figure*}

Variability of the infrared flux of NGC 2992 was already reported by Glass 
(\cite{glass97}). Apart from an outburst in 1988, which was especially strong 
in the K- and L-band, he reported a fading of NGC 2992 by about 20\% in 
  the K-band, 50\% in the L-band between 1978 and 1996, and by a factor of 
  $\approx 20$ in X-rays at the same time. 
Then, in 1998 BeppoSAX observations showed the X-ray  
flux to be again similar to that of 1978 (Gilli et al. \cite{gilli00}). Also 
line properties of Balmer- and Paschen-lines in the near infrared were similar 
to those observed before the decline. It is obvious from Table \ref{linflux} 
that our line fluxes from 2005 (determined for a similar aperture size) are 
higher than those of Gilli et  
al. (\cite{gilli00}) from January 1999. Unfortunately we have no information 
about H$\alpha$ and Pa$\beta$ to compare them to former observations 
given in Gilli et al. 
 
Oliva et al. (\cite{oliva95}) show that the $^{12}$CO(2-0) at 
  2.2\,$\mu$m in the K-band, and the $^{12}$CO(6-3) bandhead at 1.6\,$\mu$m 
  in the H-band, can be used to separate stellar and non-stellar continua in  
AGN. However, doing so requires high signal-to-noise data in both 
bands. 
Instead, we follow the method of Davies et al. (\cite{davies07}) which uses
just the 
$^{12}$CO(2-0) bandhead to calculate the fraction of the nuclear flux   
that is stellar. By assuming that the stellar equivalent width in the nuclear 
region is about 12\,\AA, regardless of the star formation history, as 
predicted by the models created with the stellar synthesis code STARS 
(Sternberg \cite{sternberg98}; Sternberg et al. \cite{sternberg03}), which is 
comparable to Starburst99 (Leitherer et al. \cite{leitherer99}; Vazquez \& 
Leitherer \cite{vazquez05}), we calculated the non-stellar dilution using the 
equation:  
\begin{equation} 
1-D=\frac{EW_{\rm obs}}{EW_{\rm intr}} 
\end{equation}  
where $D$ is the fraction of the continuum that is non-stellar. We found an 
equivalent width of (5.0$\pm$1.6)\,\AA\ in the central $0\farcs5\times 
  0\farcs5$, implying that  
58\% of the K-band luminosity in this region comes from non-stellar components.  
With this value we determined a stellar luminosity of 
L$_{\mathrm{K}}$=3.9$\cdot$10$^7$ L$_{\sun}$ and a total continuum luminosity 
  of L$_{\mathrm{K}}^{\mathrm{cont}}$=5.4$\cdot$10$^7$ L$_{\sun}$ in this 
  central region in the K-band.

   \begin{figure*} 
   \centering 
   \includegraphics[width=16cm]{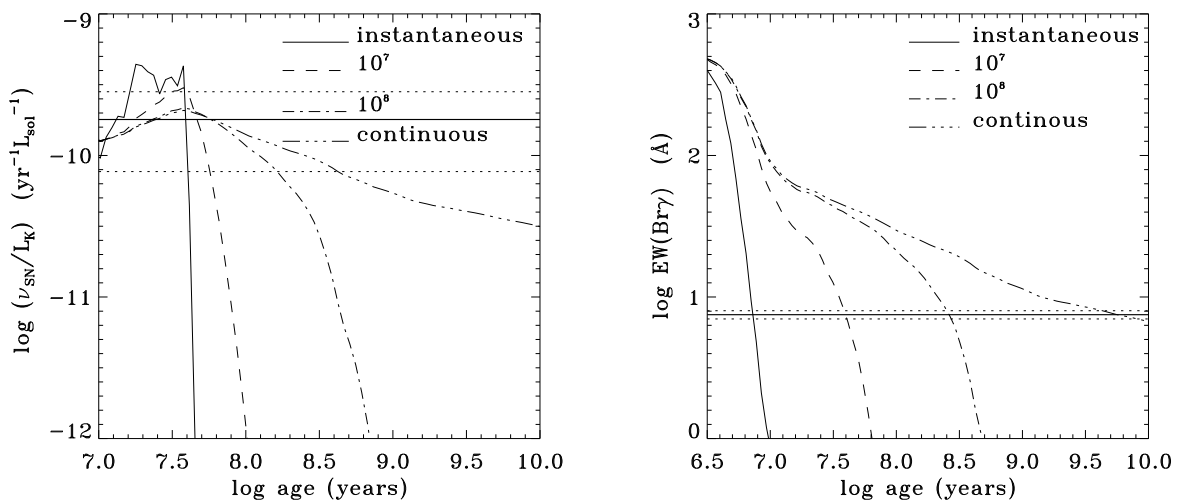} 
      \caption{Supernova rate and Br$\gamma$ equivalent width calculated with 
        STARS as a function of age for several star formation 
        timescales. They are normalised to the K-band stellar continuum and 
        L$_K$ is the total luminosity in the 1.9--2.5\,$\mu$m band  
        in units of the solar luminosity (L$_{bol}$=3.8$\times10^{26}$W). The 
        observed values are drawn as horizontal lines, the error margin is 
        indicated by dashed lines.} 
         \label{stars} 
   \end{figure*}

\subsection{2-dimensional flux distributions, velocities, and velocity dispersions} 
2-dimensional flux distributions, velocities, and velocity dispersions of individual 
lines were determined by fitting the unresolved line profile of an OH line, 
which was convolved with a Gaussian, to the respective lines. 
The continuum level was found by a linear fit to the spectrum and subtracted. 
The uncertainty of a fit was estimated using $\chi^2$ techniques by 
refitting the best-fit Gaussian with added noise at the same level as the 
data 100 times. Standard deviations of the best-fit parameters are then used 
as the uncertainties. Uncertainties of velocities and velocity dispersions are 
of the order of 10\,km\,s$^{-1}$ to 15\,km\,s$^{-1}$ for the $^{12}$CO(2-0) 
bandhead at 2.2\,$\mu$m (i.e. for the stars) and 
Br$\gamma$, and about 5\,km\,s$^{-1}$ for H$_2$\,1-0S(1). Errors of the 
continuum range from below 1\% for Br$\gamma$ and H$_2$\,1-0S(1), and 6\% for 
the stars.  
 
In Fig. \ref{kine} we show the morphology of the stellar continuum, the 
Br$\gamma$- and H$_2$\,1-0S(1) lines and their  
respective velocities and velocity dispersions. The spatial resolution is 0.3\arcsec 
corresponding to a length scale of about 50\,pc; the spectral resolution is 
83\,km\,s$^{-1}$. The continuum seems to trace an inclined disk with the 
north-west side more obscured. Given the broad dust lane seen in optical 
images this is not surprising.  
 
There exist remarkable differences. Firstly, the morphology of the Br$\gamma$ 
line does not follow the stellar continuum. To make this more obvious the 
contour lines of the Br$\gamma$ continuum are overlaid in Fig. \ref{kine} on 
the stellar continuum. The center of the stellar continuum is shifted to the 
south east and the intensity distribution is elongated in the north east 
to south west direction in contrast to the circular Br$\gamma$ contour. This 
means that part of the Br$\gamma$ emission must be attributed to other sources 
than a starburst. On the other side, the morphology of the H$_2$\,1-0S(1) line 
is similar to the stellar continuum. 
 
The zero line of the velocity is orientated roughly east west in all three 
velocity maps. The maximum velocities range from $+$100\,km\,s$^{-1}$ to 
$+$170\,km\,s$^{-1}$ and the minimum  
velocities from $-$100\,km\,s$^{-1}$ to $-$140\,km\,s$^{-1}$.  
The velocity map of the stars shows a distinct 
increase to up to 100\,km\,s$^{-1}$ at the position of the centre of the 
Br$\gamma$ continuum.   
 
The stellar component, Br$\gamma$ and H$_2$\,1-0S(1) all show velocities 
away from us in the southwest and towards us in the northeast. The velocity 
field of H$_2$\,1-0S(1) and also that of Br$\gamma$, albeit not as clearly, show 
redshifted emission also to the southeast. Garc\'ia-Lorenzo et 
al. (\cite{garcia01}) investigated the velocity field of [\ion{O}{iii}] and 
found an arc of redshifted velocities to the southeast and one of blueshifted 
velocities to the northwest with velocities of ($64\pm18$)\,km\,s$^{-1}$, and 
($-79\pm28$)\,km\,s$^{-1}$, respectively. These arcs are just outside our 
field of view. Only the redshifted velocities in the south east corner of 
H$_2$ might be caused by them (Fig. \ref{radio}). Since these ``arcs'' 
resemble the loops of the figure-of-8 when  
superimposed on the radio emission there might be a relation 
between both structures. However, this would suggest an inflow, if the 
southeast part of the figure-of-8 is closer to us, as generally 
assumed. In the picture of expanding gas bubbles this cannot be understood. 
  
Already Chapman et al. (\cite{chapman00}) proposed that the radio 
morphology consists of a component out of the plane of the galaxy disk and one 
within the disk associated with a spiral arm. On the basis of the narrow and 
well defined spectral lines within the arcs Garc\'ia-Lorenzo et al. conclude 
that the arcs are due to gas in the disk rather than in the loops. However, 
none of the arcs is associated with the structures that Chapman et 
al. interpret as spiral arms. 
 
There is a small region in the velocity field of Br$\gamma$ which extends from 
west of the centre to nearly the edge of the FOV in the SW with a lower  
velocity relative to its surroundings. This can be interpreted  
as an outflow superimposed on the galaxy rotation. Although it coincides with 
the southern inner spiral arm found  
by Chapman et al. (\cite{chapman00}) in their model-subtracted R- and H-band 
images, it cannot be related to it because then the spiral arm would move 
relative to the ambient medium. 
There is no counterpart in the H$_2$\,1-0S(1) and stellar continuum velocity 
fields. 
 
Common to both the stars and the lines are relatively  
high velocity dispersions of up to 250\,km\,s$^{-1}$ for the stars and 
Br$\gamma$, and 130\,km\,s$^{-1}$ for H$_2$\,1-0S(1). However, the stellar  
dispersion map shows no structures, except that there seems to be a  
local minimum at the position of the centre of the Br$\gamma$ 
continumm. There, the  
measured velocities range from $(110-150)$\,km\,s$^{-1}$ $\pm 
(20-30)$\,km\,s$^{-1}$ compared to the 
surrounding values of more than $(200\pm15)$\,km\,s$^{-1}$. This high stellar 
velocity dispersion which we observe even on the smallest scales suggests that 
the K-band light is dominated by stars in the bulge.  
 
The high velocity dispersion of Br$\gamma$ is confined to the inner 1\farcs2, 
outside dropping to $(0- 30)$\,km\,s$^{-1}$. The dispersion of 
H$_2$\,1-0S(1) shows maxima  
of 150\,km\,s$^{-1}$ to the southwest and northeast of the  
centre, which correspond with the velocity maxima and minima. They also lie at 
the edge of radio features within the loops (Fig.\ \ref{radio}) and may be 
associated with  
inner spiral arms. The dispersion of H$_2$\,1-0S(1) also  
shows a local minimum at the centre where the dispersion drops to 
70\,km\,s$^{-1}$. There might also be a local maximum at the position of the 
centre of the Br$\gamma$ contour.  
 
The complexity and the differences of 
velocity and dispersion maps of the stars and the lines might be due  
to the superposition of the galaxy rotation and an outflow as it was already 
suggested by several authors (e.g. Chapman et al. \cite{chapman00}).

\section{Constraining nuclear star formation in NGC 2992} 
 
The possible existence of a nuclear starburst in NGC 2992 was 
investigated by Davies et al. (\cite{davies07}), who compared the 
radial profile of  
  the stellar continuum to an $r^{1/4}$ law and an exponential 
  profile.  
They found that there might be excess continuum at 
  $r<0\farcs5$ due to a distinct stellar population superimposed on 
  the bulge.  
However, although the size scale matched the nuclear star forming regions in
other AGN with typical HWHM  
  of 10\,pc--50\,pc (s. Fig. 5 in Davies et al. \cite{davies07}), they 
  did not reach a conclusion on NGC 2992 because the excess was only 
  apparent for an exponential profile. 
Data at a higher resolution of about 0\farcs1 would be needed to 
explore the spatial properties further. 
Instead, here we look at various star formation diagnostics in order 
to assess the luminosity and age of a putative nuclear starburst. 
 
The stellar population synthesis code STARS (e.g. Sternberg 
\cite{sternberg98}; Sternberg et al. \cite{sternberg03}) was also used to model the 
equivalent width of Br$\gamma$ and the supernova rate. The program follows the 
evolution of a cluster of stars through the Hertzsprung-Russell-Diagram depending on 
the star formation history. We assumed decaying star 
formation rates with timescales of $10^6$ (instantaneous), $10^7$, $10^8$, and 
$10^{12}$ (continuous) years. In Fig. \ref{stars} the results from STARS 
together with the observations are presented. Further details on star 
formation diagnostics can be found in Davies et al. (\cite{davies07}).

\subsection{Br$\gamma$ equivalent width and supernova rate} 
Using the stellar continuum luminosity from section 3.1, an 
upper limit for the equivalent width of Br$\gamma$ associated with star 
formation from the narrow Br$\gamma$ line flux can be estimated. However, for 
NGC 2992 it is  
difficult to quantify what fraction of Br$\gamma$ is associated with star  
formation, since the morphology of the line does not follow the 
stars (Fig. \ref{kine}). Especially the south-west side shows velocities which 
are blue shifted relative to their environment indicative of motions towards us. 
The different morphology of H$_2$\,1-0S(1) and  Br$\gamma$ found in section 3.2 points 
towards a kinematically different origin, and supports the conclusion that at 
least part of the Br$\gamma$ emission is due  
to the AGN and not to star formation. Thus the narrow Br$\gamma$ line flux 
includes a certain unknown fraction from  
the AGN and the derived equivalent width of $(7.5\pm0.5)$\,\AA\ is only an 
upper limit.  
This value, and all the numbers throughout the rest of the paper, 
are determined in the central $0\farcs5\times0\farcs5$ unless 
otherwise specified.
 
We estimate a supernova rate  
for NGC 2992 from the unresolved radio flux of 7\,mJy at 5 GHz (Wehrle 
\& Morris \cite{wehrle88}) and the estimation done by Sadler et 
al. (\cite{sadler95}) that the core flux at 5\,GHz is less than 6\,mJy. Their 
estimation is based 
on their higher spatial resolution measurements at 2.3\,GHz and the non-detections 
at 1.7\,GHz and 8.4\,GHz. Taking a flat spectral index (Chapman 
et al. \cite{chapman00}) the 5\,GHz flux will not be much less than 6\,mJy, 
which would result in about 1\,mJy extended emission. If we assume that it 
can be attributed to star formation, a supernova rate of 0.003 yr$^{-1}$ can 
be derived with the relation of Condon (\cite{condon92}). On the other hand, 
if we assume a spectral index of 0.82, which was derived by Ulvestad \& Wilson   
(\cite{ulvestad84}) for the unresolved flux at 1.5\,GHz and 5\,GHz, together 
with the 2.3\,GHz flux of Sadler et al. (\cite{sadler95}) we get a 
flux of about 3\,mJy at 5\,GHz. This leaves us with about 4\,mJy 
extended emission corresponding to a supernova rate of 0.011 yr$^{-1}$. We 
consider the derived supernova rates as extremes and adopt a mean value of 
0.007 yr$^{-1}$.  
 
We can now try to determine a star formation age from the observed Br$\gamma$ 
equivalent width and the supernova rate by comparing them to the theoretical 
values calculated by STARS, which are plotted in Fig. \ref{stars} for 
different star formation timescales. It is immediately obvious that for a continous 
star formation the observed Br$\gamma$ equivalent width and the supernova 
rate yield inconsistent star formation ages of more than 3 Gyr and (25 -- 
50)\,Myr, respectively. In addition, an ongoing star formation would require 
an equivalent width of Br$\gamma$ of 10\,\AA\ -- 15\,\AA , which is not 
observed.  
 
An intermediate decay timescale of $10^8$\,yr yields a star formation 
age from the Br$\gamma$ equivalent width of about 280\,Myr and 25\,Myr -- 
60\,Myr from the supernova rate, which is an order of magnitude lower. Even if 
we consider the lower boundary of the supernova rate of 0.003\,yr$^{-1}$ the 
resulting star formation age of 160 Myr is well below the age derived from the 
Br$\gamma$ equivalent width.  
 
The best agreement between the derived star formation ages from the Br$\gamma$ 
equivalent width and the supernova rate can be achieved for a decay timescale 
of $10^7$\,yr. The resulting star formation ages amount to 40\,Myr and 
50\,Myr respectively. 
The largest error of $\pm$ 10\,Myr is due to the uncertain radio flux at 5\,GHz and 
subsequently the supernova rate. The unknown contribution of the AGN to 
Br$\gamma$ does not contribute significantly to the uncertainty, because the 
curve is steep. Even an increasing contribution of the AGN up to nearly 
  100\%, which leads to a respective lower stellar contribution to the 
  Br$\gamma$ equivalent width, would  
  increase the age only up to 60\,Myr. This is still within the errors set by 
the supernova rate. We therefore conclude that a stellar population exists in 
NGC 2992 with an age of 40\,Myr to 50\,Myr.  
 
In principal 
there is also agreement between the Br$\gamma$ equivalent width and the 
supernova rate for instantaneous star formation at a very young age of about 
10\,Myr, but we consider this less likely; partly because the timescales are much  
shorter, but primarily because it requires very careful tuning to 
match the increasing supernova rate and the quickly decaying 
Br$\gamma$ equivalent width.   
 
Combining the age range of 40\,Myr -- 50\,Myr with the bolometric luminosity 
and black hole mass for NGC 2992 allows us to locate this galaxy on 
Figure~11 of Davies et al. (\cite{davies07}). 
This figure shows how the 
luminosity of an AGN might be related to the age of the starburst, and 
NGC 2992 can be placed in the lower left where the starburst age is 
young and the AGN is, at the present time, accreting at a rate that is 
low compared to the Eddington rate. 
The galaxy thus appears to fit into the proposed scheme, where 
there is either a delay between the onset of starburst activity and the onset 
of AGN activity, or star formation ceased once the black hole become active 
(s. Davies et al. \cite{davies07} for a detailed discussion). 
 
We also briefly consider whether the interaction between NGC\,2992 and 
NGC\,2993 might be responsible for the starburst. 
The gravitational (tidal) force during an interaction can be 
  approximated as a delta-function that occurs at perigalacticon. 
According to Duc et al. (\cite{duc00}) perigalacticon of NGC 
  2992 and NGC 2993 occurred about 100\,Myr ago.  
At that time the gas would have felt the strongest impulse.  
Given that it takes time for the gas to respond 
  and, for example, be driven to the centre, it is  
  perhaps plausible for the interaction to have triggered the starburst. 
 
\subsection{Mass-to-light-ratio and bolometric luminosity} 
In order to estimate the mass-to-light-ratio we determined the dynamical mass 
within 0\farcs5 from the following relation: 
\begin{equation} 
M_{\rm dyn}=(v_{\rm rot}^2 + 3\sigma^2)\cdot R / G 
\end{equation} 
where $v_{\rm rot}$ is the observed velocity corrected by the inclination and 
$\sigma$ is the observed velocity dispersion at radius $R$ (Bender et 
al. \cite{bender92}; Davies et al. \cite{davies07}). We did not correct for 
the molecular gass mass nor for the contribution of an old stellar 
population. Therefore the derived dynamical mass of 8$\cdot$10$^7$\, M$_{\sun}$ 
is only an upper limit. With a black hole mass of 5$\cdot$10$^7$\, M$_{\sun}$ 
determined from velocity dispersion measurements (Woo \& Urry 
\cite{woo02}) the dynamical mass is at most 3$\cdot$10$^7$\, M$_{\sun}$ for  
the young population. Taken at face value, the resulting
  mass-to-light-ratio of 0.8 speaks in 
  favour of a star formation age greater than 80\,Myr for all star 
formation rates. This is a factor of 2 higher than the age derived from the 
Br$\gamma$ equivalent width and the supernova rate. However, it needs 
to be borne in mind that the dynamical mass 
is only an upper limit of the younger stellar population (it will 
contain a significant contribution from the older bulge population) and the 
  uncertainty of the black hole mass can be nearly as high as 50\% 
  (s. Woo \& Urry \cite{woo02}).  
Either a change of the black hole mass by 10\% or a reduction of 
the mass of the young population by 20\% would bring the star 
formation ages into agreement.  
 
STARS also allows us to estimate the stellar bolometric luminosity. The ratio 
between $L_{\rm bol}$ and $L_{\rm K}$ depends on the age and the exponential decay 
timescale of the star formation. Given the star formation age of 40\,Myr -- 
50\,Myr we derive a bolometric luminosity for NGC 2992 of  
70--100 times the K-band luminosity or (3--4)$\cdot$10$^9$ L$_{\sun}$, which 
is (6--8)\% of the total bolometric luminosity of NGC~2992 
(5$\cdot$10$^{10}$ L$_{\sun}$ calculated between 8 and 1000\,$\mu$m from 
IRAS $12-100\,\mu$m flux densities, and additionally taking into account the 
emission at shorter wavelengths; Davies et al. \cite{davies07}).

\subsection{Star formation rate} 
 
Using the star formation history (decay timescale and age) above, we can 
estimate the star formation rate. 
The first step is to determine what fraction of the K-band 
luminosity in the central 0\farcs5 comes from young stars. 
To do so, we estimate the contribution from an old population, by 
assuming that the dynamical 
mass is dominated by this old population. 
For an adopted age of 5\,Gyr, STARS yields a mass-to-light ratio of 
about 30\,M$_{\sun}/L_K$. 
The dynamical mass determined above of $3\cdot 10^7$\,M$_{\sun}$ then 
results in a luminosity of  
$1\cdot 10^6$\,L$_{\sun}$ for the old population, one order of magnitude below 
the observed K-band luminosity. Therefore the K-band luminosity is dominated by 
the young population. 
We can therefore scale the luminosity of the starburst model to the 
total K-band luminosity in the central 0\farcs5, yielding an initial 
star formation rate of 4.3\,M$_{\sun}$yr$^{-1}$ for an age of 
50\,Myr. 
Since the SFR is decaying, an alternative, perhaps more meaningful, number may be the 
time-averaged star formation rate -- which we define as simply the mass of stars 
formed divided by the age. This yields 
$\left\langle\rm{SFR}\right\rangle\sim1$\,M$_{\sun}$yr$^{-1}$. 
While this appears modest, in terms of SFR per unit area it corresponds 
to $\sim200$\,M$_{\sun}$yr$^{-1}$kpc$^{-2}$, within the range of 
values found for the nuclear starbursts in other nearby AGN by Davies 
et al. (\cite{davies07}).

An independent method to estimate the SFR is via the polycyclic aromatic 
hydrocarbon (PAH) features. 
Farrah et al. (\cite{farrah07}) employed diagnostics based on the luminosities 
and equivalent widths of fine-structure emission lines and PAH 
features as well as the strength of the 9.7\,$\mu$m silicate  
absorption feature to investigate the power source of the infrared emission in 
ULIRGs. With these tools they were able to distinguish between starburst and 
AGN caused infrared emission, and to determine a star formation rate. 
 
In order to apply these diagnostics to NGC 2992 we used near infrared pipeline 
reduced IRS spectra from the Spitzer archive. If more than one spectrum 
was available we took the mean value from all spectra to determine 
line luminosities. 25\,$\mu$m and 60\,$\mu$m fluxes were taken from the IRAS 
Revised Bright Galaxy Sample  
(Surace et al. \cite{surace04}). The results are given in Table \ref{pah}.  
 
The star formation rate 
\begin{equation} 
SFR \ \left[\rm M_{\sun}\, yr^{-1}\right]=1.18\times10^{-41}(L_{\rm 6.2\mu 
  m}+L_{\rm 11.2\mu 
  m})\left[\rm erg\, s^{-1}\right] 
\end{equation}  
was empirically derived from the PAH 6.2\,$\mu$m + 11.2\,$\mu$m luminosity for 
a star formation occurring on timescales of $10^7 - 10^8$ years (Farah et 
al. \cite{farrah07}). It yields 2.5\,M$_{\sun}$ per year or about  
10\,M$_{\sun}$yr$^{-1}$kpc$^{-2}$, if we take the slit width of 3\farcs6 of the 
IRS spectra as the size of the investigated region for NGC 2992. 
This value is consistent with the SFR derived above, being above the 
time averaged rate but below the initial peak rate. 
While it can be regarded as fairly normal compared to the star  
formation rates of 100 M$_{\sun}$ per year for nearby luminous 
starbursts (e.g. Weedman \& Houck \cite{weedman08}), in terms of SFR 
per unit area, it is as extreme as these luminous starbursts.

\subsection{Starformation versus AGN contribution} 
The strength of the 7.7\,$\mu$m PAH feature (1.7$\pm$0.02) indicates that the
AGN in NGC 2992 contributes more than 50\% to the infrared flux between
$1\,\mu$m and $1000\,\mu$m (Genzel et al. \cite{genzel98}).  
[\ion{Ne}{V}] at 14.32\,$\mu$m, which is strong in AGN spectra but not in star 
forming regions, and [\ion{O}{iv}], which is strong in star forming regions, 
are both present in the spectra of NGC 2992, pointing to a co-existing 
starburst and AGN activity. 
 
In order to quantify this a little more we followed Farrah et 
al. (\cite{farrah07}) and compared the equivalent width of the PAH at 
6.2\,$\mu$m (EW=$(0.084\pm 0.002)$\,$\mu$m) to the ratio of  
[\ion{Ne}{v}] to [\ion{Ne}{ii}] and [\ion{O}{iv}] to [\ion{Ne}{ii}], 
respectively. From these diagnostics we derive a 10\% -- 20\% starburst 
fraction and about 50\% AGN fraction to 
  the infrared flux for NGC 2992 in agreement with the result  
from the PAH 7.7\,$\mu$m feature. The AGN  
contribution can also be estimated from the ratio of the 25\,$\mu$m flux to the 
60\,$\mu$m flux against the [\ion{Ne}{v}] to [\ion{Ne}{ii}] ratio. Again the 
AGN contribution amounts to about 50\%. 
 
In this and the previous section we utilized different methods to estimate the 
  contribution of stellar, non-stellar and starburst components to the K-band 
  and infrared flux (1\,$\mu$m -- 1000\,$\mu$m). Independent of wavelength 
  range and FOV we found a 50\%  
  -- 60\% AGN contribution. From the CO bandhead at 2.2\,$\mu$m we found that 42\% 
of the K-band flux in the central $0\farcs5\times0\farcs5$ can be attributed 
to stars and from the star formation history that these stars must be 
young. Also in the $3\arcsec\times3\arcsec$ FOV the K-band light  
is dominated by stars. However, according to the high stellar velocity 
dispersion it is more 
likely that stars from the bulge dominate in this larger FOV. Finally, 
  we 
find from PAH line diagnostics in the infrared a 10\% -- 20\% contribution to 
the infrared flux from a starburst. 
In addition to the AGN and starburst components, there 
still remains about 20\% -- 40\%, which must be attributed to a third component 
-- most likely the disk and bulge of the galaxy. This means that while the 
young starbust is important in the central tens of parsecs, it is still only a 
small  part of the integrated luminosity of the galaxy. Therefore we 
conclude that in NGC 2992 we see a dominating AGN activity with a relatively 
minor contribution from a starburst. 
 
\begin{table}[t] 
\caption{Measured line luminosities of fine-structure 
  lines and PAHs for NGC 2992}             % title of Table 
\label{pah}      % is used to refer this table in the text 
\centering                          % used for centering table 
\begin{tabular}{l r r}        % centered columns (4 columns) 
\hline\hline                 % inserts double horizontal lines 
Line & $\lambda^a$ & luminosity\\      
     &$$($\mu$m)$$& (10$^{33}$ W)\\ 
\hline                        % inserts single horizontal line 
PAH&6.2&9.2$\pm$0.7\\ 
PAH&11.2&12.5$\pm$0.2\\ 
$\left[\ion{S}{iv}\right]$&10.5&2.0$\pm$0.01\\ 
$\left[\ion{S}{iii}\right]$&18.7&3.8$\pm$0.3\\ 
$\left[\ion{Ne}{ii}\right]$&12.8&6.8$\pm$0.9\\ 
$\left[\ion{Ne}{iii}\right]$&15.6&7.1$\pm$0.4\\ 
$\left[\ion{Ne}{v}\right]$&14.3&3.9$\pm$1.1\\ 
$\left[\ion{O}{iv}\right]$&25.9&14.7$\pm$1.6\\ 
\hline                                   %inserts single line 
\end{tabular} 
 
$^a$ wavelengths are in the rest frame 
\end{table}

\section{Origin of nuclear outflow} 
\subsection{Starburst or AGN driven outflow} 
Heckman et al. (\cite{heckman90}) provided evidence that galaxies with a  
central starburst have large scale mass outflows which are presumably 
driven by the kinetic energy supplied by supernovae and winds from massive 
stars. In order to examine whether the outflow in NGC 2992 can be caused by 
the starburst we calculated the radius $r$ in kpc of a bubble which is 
inflated by  
energy injected at a constant rate and expanding into a uniform medium with 
a density $n_0$ in cm$^{-3}$ (Heckman et al. \cite{heckman90}). From 
\begin{equation} 
dE/dt \approx 3\times 10^{41} r^2_{\rm kpc} v^3_{\rm 100} n_0\ {\rm ergs\ s^{-1}} 
\end{equation} 
with the velocity (in 100\,km\,s$^{-1}$) $v_{\rm 100}\approx 2$ (Allen et 
al. \cite{allen99}), $n_0 \approx 3$\,cm$^{-3}$ (Heckman et al. \cite{heckman90}),  
and $dE/dt$ equal to the mechanical luminosity from stellar winds and SNe 
taken from starburst99 for instantaneous ($10^{40}$\,erg\,s$^{-1}$) star formation we get 
a radius of 40\,pc. This radius is too small compared 
to the observed radii of about 600\,pc of the figure-of-8 and 2\,kpc for the 
outflows found in H$\alpha$ and [\ion{O}{iii}] (Allen et 
al. \cite{allen99}). An energy injection of $3\cdot 10^{43}$\,erg\,s$^{-1}$ would be 
required to explain the observed radius and velocity of the outflow in 
NGC 2992. This estimation is based on a symmetric geometry. If we 
instead assume a cone with an opening angle of $116\degr$, as seen for 
the large scale outflow (Allen et al. \cite{allen99}), the energy needed is a  
factor of 10 lower but still cannot be provided by the starburst.  
 
The 2\,kpc extension of the large scale outflow and the velocity of 
200\,kms$^{-1}$ suggest  
that the outflow originated in an event at most 10 Myr ago. This is 
inconsistant with our derived star formation age of (40 -- 50)\,Myr but in 
agreement with the typical timescale of the active phase  
of a black hole in an AGN. We note that although it is also consistent 
with the age for instantaneous star formation, this scenario we 
regarded as rather unlikely since it requires fine-tuning (see 
section 4.1). 
 
Thompson et al. (\cite{thompson05}) showed that optically thick pressure 
supported starburst disks have a characteristic flux of about 
$10^{13}$\,L$_{\sun}$kpc$^{-2}$. Above this limit matter would be blown 
away. With the derived luminosity of $(3-4)\cdot  
10^9$\,L$_{\sun}$ within the central 0\farcs5 we estimated the characteristic 
flux for NGC 2992 to be $(5-7)\cdot 10^{11}$\,L$_{\sun}$kpc$^{-2}$ which is well 
below this limit. Only when the starburst was at its peak luminosity, 
which for our preferred model is nearly 10 times greater than 
the current luminosity, would the star formation approach -- although 
remaining below -- this threshold.  
The conclusion from both energy and timescale considerations, 
is that for the starburst to drive the outflows it would have to be more 
intense and more recent -- both of which are ruled out by our 
analysis. 
 
Instead, we consider whether the outflows in NGC 2992 can be driven 
by the AGN.  
In order to estimate the AGN luminosity of NGC 2992 we use the spectral energy 
  distribution (SED) of NGC 1068 as a template.  
For this SED, the optical (750\,nm) to 100\,keV luminosity is a factor 
  11.8 greater than that in the 20\,keV--100\,keV range (Pier et 
  al. \cite{pier94}). 
For NGC 2992, the luminosity in this latter range is  
$(1-4)\cdot 10^{43}$\,erg\,s$^{-1}$ (Beckmann et 
  al. \cite{beckmann07}). 
Thus the total luminosity of NGC 2992 is about an order of magnitude 
  greater, which yields about $10^{44}$\,erg\,s$^{-1}$.  
This is a factor of 3--30 times greater than the energy estimated above that
is required to generate the outflow.
We therefore conclude that it is the AGN, rather than the starburst, 
  that has driven the outflow.

\subsection{Velocities along the major axis} 
\begin{figure} 
   \centering 
   \includegraphics[width=8truecm]{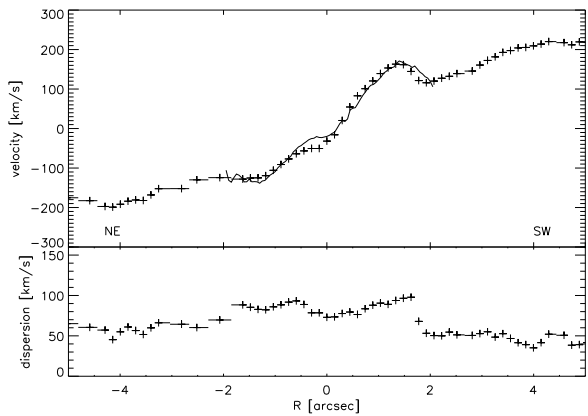} 
      \caption{Radial velocities (top) and corresponding velocity dispersion 
        (bottom) from longslit spectra taken with ISAAC along  
        the major axis of NGC 2992 (crosses). Overplotted is a radial velocity 
        curve drawn from SINFONI data also along the major axis. 1\arcsec\ 
        correspond to 150\,pc, northeast is to the left, southwest to the 
        right. } 
         \label{isaac} 
   \end{figure}

Longslit spectra were taken with ISAAC along the 
major axis (position angle PA = 34$\degr$) at 2.1\,$\mu$m.   
The resulting velocity curve of H$_2$\,1-0S(1) along the slit shows 
a steep increase to the southwest (decline to NE) up to a radius of 1\farcs5 
away from the centre followed by a sharp decline within about 0\farcs5 and 
subsequently a shallower incline. To the northeast the velocity curve behaves 
similar. This is also observed in the SINFONI data (Fig. \ref{isaac}), and 
is in agreement with published data of Marquez et 
al. (\cite{marquez98}) for a position angle of 30\degr\ and Veilleux et 
al. (\cite{veilleux01}), too. 
 
Colina et al. (\cite{colina87}) found asymmetric line profiles for a 
region with $r\le 3$\arcsec, and symmetric profiles for $r\ge 3$\arcsec. They 
interpreted their results as dynamically decoupled 
nuclear and off-nuclear regions. Their velocity curve of [\ion{O}{iii}] is in 
agreement with our velocity curve. However, due to their  
coarse sampling along the major axis they could not resolve less than 
3\arcsec. 
 
\begin{figure} 
   \centering 
   \includegraphics[width=8.2truecm]{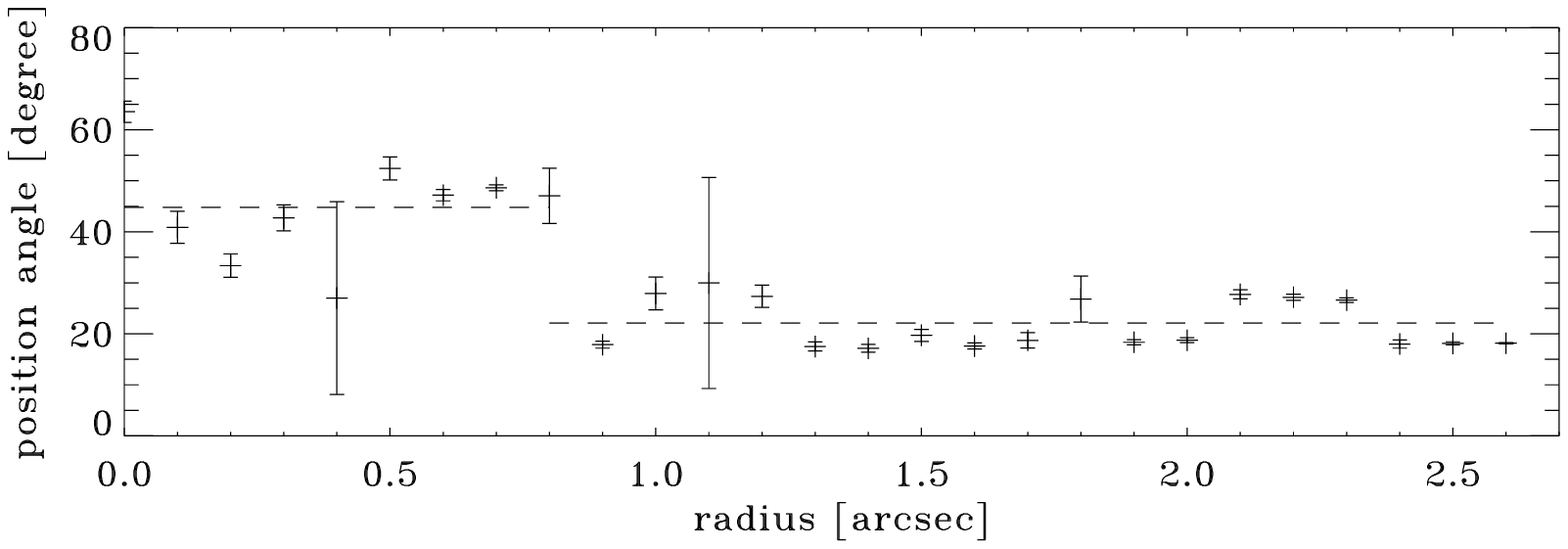} 
   \includegraphics[width=8.2truecm]{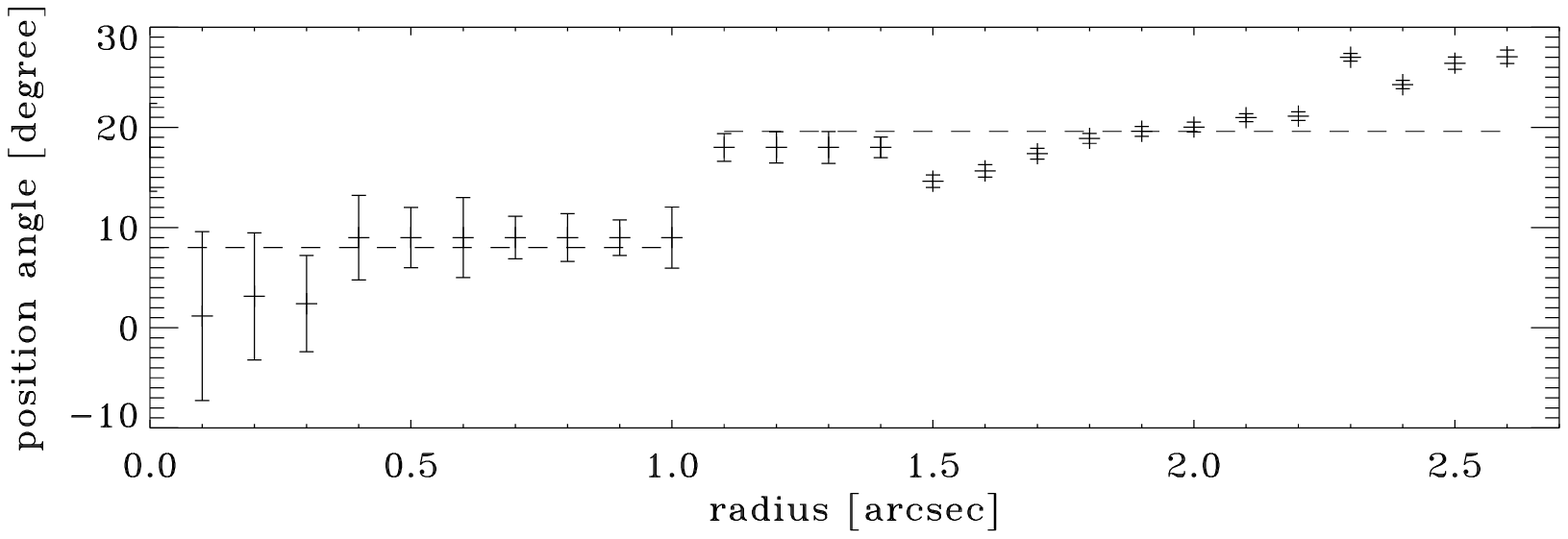} 
      \caption{The dependence of the position angle on the distance from the 
        centre for the 2-dimensional stellar (top) and H$_2$\,1-0S(1) 
        data (bottom). The mean values of the PA for radii below and above 
        1\arcsec\ are also given (dashed line).} 
         \label{pa} 
   \end{figure}

Evidence for dynamical decoupled nuclear and off-nuclear regions comes also 
from kinemetry (Krajnovi\`c et  
al. \cite{krajnovic06}) of our 2-dimensional stellar velocity data. With the 
position  
angle and inclination as free parameters the PA changes from $45\degr\pm 
11\degr$ to $22\degr\pm 5\degr$ at about 1\arcsec\ (Fig. \ref{pa}). This is 
in good agreement  
with the 22.5\degr\ of Jarrett et al. (\cite{jarrett03}) derived from 
the K$_S$-band 20 mag\,arcsec$^{-2}$ isophot. The PA of 
45\degr\ better fits to the inclination of the accretion disk of about 
37\degr\ and  
46\degr, derived by Gilli et al. (\cite{gilli00}) with the assumption of both 
Schwarzschild and Kerr metrics from disk line models and iron line profiles,  
 than to the 34\degr\ of Garc\'ia-Lorenzo et 
al. (\cite{garcia01}). The PA of the 2-dimensional H$_2$\,1-0S(1) velocity data 
also changes at 1\arcsec, however from $8\degr\pm4\degr$ to 
$20\degr\pm4\degr$ (Fig. \ref{pa}). That the changes in PA are derived 
  from kinematic rather than photometric data, argues that they 
  are real and not just an effect of extinction. 
It therefore seems that the stars and the gas show a  
different behaviour in the very centre and a similar one outside.     
 
\subsection{A geometric model for the inner $3\,\arcsec\times3\,\arcsec$} 
In order to investigate the different components contributing to the 
velocity field and to explain the different position angles observed in the 
central $3\,\arcsec\times3\,\arcsec$ of NGC 2992 
we developed a simple geometric model. It consists of an inner and an outer 
disk, in which particles were assumed to be in circular motion in a 
gravitational potential. The radius of the inner disk is determined by the 
  change in PA found in kinemetry of stellar and H$_2$\,1-0S(1) data at
  about 1\arcsec. It is 
  not related to the starburst nor does it reflect the sphere of gravitational 
  influence of the black hole, which is only a few parsec in radius. The radius 
  of the outer disk is cut at 
  5\arcsec\ due to our FOV of 3\arcsec. In addition, a conical outflow with 
constant velocity represents the figure-of-8. All components could  
be rotated by arbitrary angles around the x-, y-, and z-axis, which points to 
the observer, thus allowing for 
different orientations to each other. Velocities were adjusted in such a way, 
that they  
reflect the observed velocity curve. Due to the simplicity of the model, we 
only can draw some qualitative conclusions.  
 
There is an inner and outer region which must have different orientations in 
space (realized in the model by different rotation angles around the x-axis) 
in order to explain the different position angles. However, the  
inclination angle of 70\degr\ and the position angles cannot be modelled 
exactly at the  
same time. Dependent on the start value for the ellipticity of the unrotated 
outer disk we get an inclination angle of 50\degr\ for a circular disk and 
63\degr\ for an ellipticity of 0.7, and position angles of 13\degr\ and 
18\degr, respectively. For the inner, circular region we get an inclination 
angle of 60\degr\ and a position angle of 34\degr. The different inclination 
angles might point to a dynamical decoupled inner and outer region as it was 
already suggested by Colina et al. (\cite{colina87}). 
 
In order to achieve a position 
angle of the figure-of-8 consistent with the published $-26$\degr\ (Wherle et 
al. 1988), it has to be perpendicular to a plane whose inclination angle 
  is 51\degr, which is slightly lower than the inclination of the inner disk. One 
  can identify this plane with the plane of the accretion disk, which itself is not 
  modeled. The resulting position 
angle is $-27$\degr\ with the SE part of it pointing away from us, 
consistent with the reshifted emission there seen by Garc\'ia-Lorenzo et 
al. (\cite{garcia01}). 
 
The derived geometry from our model yields the values from the  
literature of 15$\degr$ for the position angle of the outer region 
(photometric axis), 34$\degr$ for the inner  
5\arcsec\ (Garc\'ia-Lorenzo et al. \cite{garcia01}), and the inclination of the 
accretion disk of $46\degr\pm7\degr$ (Gilli et al. \cite{gilli00}) very 
well. The resulting velocity field from this geometry is shown in 
Fig.\ \ref{veltheo} and is in good agreement with the measured velocity 
field of H$_2$\,1-0S(1) depicted in the lower panel of 
Fig.\ \ref{kine}.  
Within this geometry the hump in the velocity curve at 1\farcs5 is the 
signature of the  
cone describing the figure-of-8, which is superposed on the inner  
disk. The stellar velocity curve along the major axis from SINFONI and ISAAC 
does not show these humps at 1\farcs5 which speaks in favour of this 
interpretation. 
Our model suggests that the outer disk, inner disk, and 
  accretion disk are at different orientations. 
One possible reason for this is that this geometry reflects a single 
warped disk, or even a warped outflow. 
However, it is beyond the scope of this paper to explore all the 
options and to assess the physical reason behind the observed 
geometry.

   \begin{figure} 
   \centering 
   \includegraphics[width=8truecm]{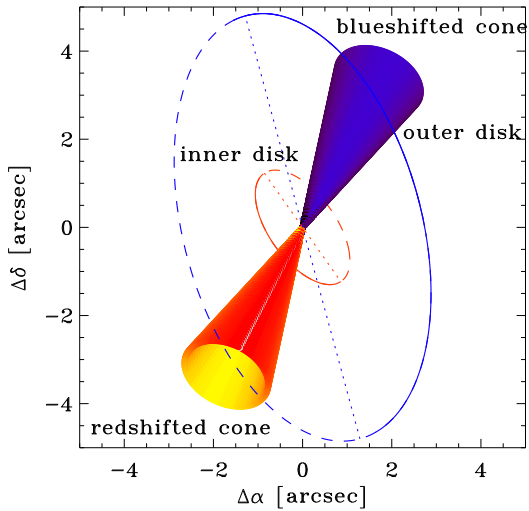} 
      \caption{Cartoon which shows the geometry of the central 
        region of NGC~2992. Those parts of the disks with solid lines point 
        towards us, those with long-dashed away from us. Short-dashed lines 
        indicate the position angles of 15\degr\ for the outer and 
        34\degr\ for the inner disk.} 
         \label{geomodel} 
   \end{figure}

   \begin{figure} 
   \centering 
   \includegraphics[width=7.truecm]{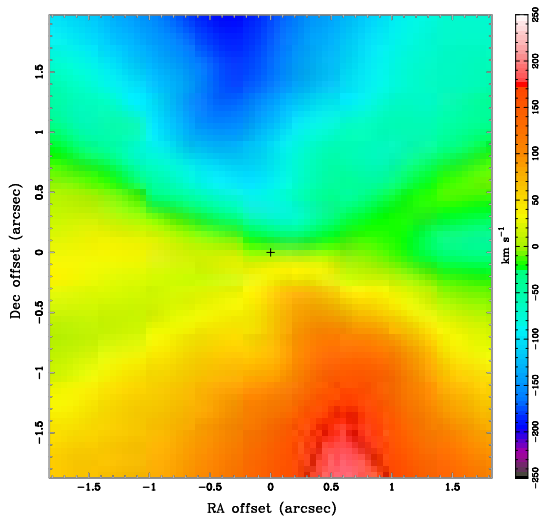} 
      \caption{Calculated velocity field in the central region of NGC 2992 
        resulting from two inclined disks and a conical outflow 
        (see subsection 5.3.).} 
         \label{veltheo} 
   \end{figure} 
 
\section{Conclusions} 
We have presented a detailed analysis of near infrared K-band spectra complemented by 
N- and Q-band spectra of NGC 2992. K-band spectra were 
obtained with the adaptive near infrared integral field spectrograph SINFONI 
and allow a reconstruction of the 2-dimensional distribution and kinematic 
of the stars and gas in the inner $3\arcsec\times 3\arcsec$ (450\,pc) at 
an angular resolution  
of 0\farcs3. The N- and Q-band data were obtained with Spitzer and  
taken from the archive. We compared the equivalent width of Br$\gamma$ and the 
supernova rate, which was derived from radio data from the literature, to 
STARS evolutionary synthesis models and 
find evidence for a short burst of star formation $(40 - 50)$\,Myr 
ago.  
 
From our near-infrared data as well as equivalent widths and luminosities of 
fine-structure emission lines and polycyclic aromatic hydrocarbon features detected 
in the N- and Q-band spectra we were able to estimate the nuclear 
star formation rate and quantify the contribution of the 
star formation to the AGN luminosity: the luminosity is dominated by    
the AGN activity with a contribution of only $10\%-20\%$ from nuclear star 
formation. This is in agreement with the conclusion that part of the Br$\gamma$ 
  emission cannot be attributed to star formation but is due to the AGN. 
 
Energy and timescale considerations let us conclude, that the starburst would  
have to be more recent and more intense to drive the figure-of-8 as well as 
the large scale (2\,kpc) outflow. On the other hand the estimation of the 
luminosity of the AGN of $10^{44}$\,erg\,s$^{-1}$ is about an order of 
magnitude higher than is necessary to drive the outflows. 
 
The observed velocity curve in H$_2$\,1-0S(1) of NGC 2992 can be described  
as the superposition of the galaxy rotation and a conical outflow with 
constant velocity. The different position angles of the inner and outer region 
can be modelled by disks with different orientations in space and thus being probably 
dynamically decoupled. However, these disks are not related to the 
  starburst. In this picture the south east cone is pointing away  
from us supporting the findings from Garc\'ia-Lorenzo (\cite{garcia01}).

\begin{acknowledgements} 
      The authors thank all those at MPE and ESO Paranal who were 
      involved in the SINFONI observations. This work is also based in part 
      on observations made with the Spitzer Space Telescope, which is operated 
      by the Jet Propulsion Laboratory, California Institute of Technology 
      under a contract with NASA.  
\end{acknowledgements}

\end{document}